\documentclass[%
 preprint,
superscriptaddress,
 amsmath,amssymb,
 aps,
prb,
]{revtex4-1}

\usepackage{graphicx}
\usepackage{dcolumn}
\usepackage{bm}
\usepackage{epsfig}
\usepackage{multirow}
\usepackage{wrapfig}

\begin{document}

\title{Positron annihilation spectroscopy for the pure and Niobium doped ZrCo$_2$Sn Heusler compound}

\author{D. Benea}
\affiliation{Faculty of Physics, Babes-Bolyai University, Kogalniceanustr 1, Cluj-Napoca, Ro-400084, Romania}
\author{A. \"Ostlin}
\affiliation{Theoretical Physics III, Center for Electronic Correlations and Magnetism, Institute of Physics, University of Augsburg, Augsburg 86135, Germany}
\author{J.A. Weber}
\affiliation{Physik-Department, Technische Universit\"at M\"unchen, James-Franck Stra\ss e, Garching, 85748, Germany}
\author{E. Burzo}
\affiliation{Faculty of Physics, Babes-Bolyai University, Kogalniceanustr 1, Cluj-Napoca, Ro-400084, Romania}
\author{L. Chioncel}
\affiliation{Theoretical Physics III, Center for Electronic Correlations and Magnetism, Institute of Physics, University of Augsburg, Augsburg 86135, Germany}
\affiliation{Augsburg Center for Innovative Technologies, University of Augsburg, Augsburg 86135, Germany}

\begin{abstract}
We perform spin-polarized two-dimensional angular correlation of annihilation radiation (2D-ACAR) calculations for the recently predicted ZrCo$_2$Sn-Weyl Heusler compound within 
the density functional theory using the generalized gradient approximation (GGA) and its extension GGA+U. 
We confirm that within the GGA+U method, a pair of Weyl-points are revealed, and that by
doping with Niobium, for the composition Nb$_{0.3}$Zr$_{0.7}$Co$_2$Sn, the Weyl points are reaching the Fermi level.
Our 2D-ACAR results indicate the existence of the Weyl points, however, within the 
present calculation, it is uncertain if the smearing at the Fermi level 
can be attributed to the positron wave function.
\end{abstract}

\maketitle

\section{Introduction}
\label{main}

Recently, Weyl fermions have been predicted in the ferromagnetic full Heusler  ZrCo$_2$Sn~\cite{wa.ve.16}. The band structure calculations (within GGA+U) show a half-metallic ferromagnetic behavior of this compound, with a magnetic moment of 2$\mu_B$, in agreement with the Slater-Pauling rule. Half-metals are characterized by a metallic electronic structure for one spin channel, whereas for the opposite spin direction the Fermi level is situated within an energy gap \cite{gr.mu.83,ka.ir.08}. The time-reversal symmetry is broken in these systems, therefore from the magnetic point of view these could be either ferro- or ferrimagnets with perfect spin-polarization at the Fermi level. Using the electronic structure calculations
Wang et al.~[\onlinecite{wa.ve.16}] found that the easy axis is oriented along the [110]
direction which was also confirmed by experimental measurements. Along this axis  
two Weyl points related to the inversion symmetry have been found. When alloyed 
with Nb, the spin-up Weyl points shift closer to the Fermi level~\cite{wa.ve.16} and consequently transport, spectroscopic properties such as the chiral anomaly, and unusual magnetoresistance are expected to occur in this compound. 

The angular correlation of annihilation radiation (ACAR) is a specific technique
within Positron Annihilation Spectroscopy (PAS) which allows to study the momentum density of electrons in solids in particular Fermi surfaces of metals and alloys. The behavior of positrons in condensed matter has been subject to an intense theoretical and experimental investigation and the use of positrons to probe electronic structure is well documented and reviewed~\cite{west.95,dugd.16}. Two main categories of PAS in solids are currently available: (i) bulk studies using fast positrons from radioactive $\beta^+$ sources \cite{Ceeh2016,Dugdale2013} and (ii) surface and near surface studies with variable energy positron beams \cite{Peng1996,Falub2001}. As measurements of the two-dimensional angular correlation of annihilation radiation (2D-ACAR) captures both low- and high-momentum components of the 
electronic states, it can provide useful information about the Dirac / Weyl states. It was recently shown in a combined experimental and theoretical study of the topological insulator Bi$_2$Te$_2$Se, that a bound positron state exists at its surface and that the theoretical calculation confirms 
the experiment, showing a significant overlap between the positron and the topological state~\cite{ca.sh.16}. That demonstrates that besides the angle-resolved photoemission spectroscopy and scanning tunneling spectroscopy, positron annihilation spectroscopy provides an equal highly surface sensitive probe for the topological states of matter.


In this paper we study the spectral function around the Fermi level for the ZrCo$_2$Sn and Nb$_{0.3}$Zr$_{0.7}$Co$_2$Sn Heusler compounds. Our results confirm the previously reported~\cite{wa.ve.16} existence of Weyl nodes above the Fermi level in the majority spin channel along the [110] direction, using the GGA+U approximation to the exchange correlation potential of the density functional theory (DFT)~\cite{jo.gu.89,kohn.99,jone.15}. We show that doping ZrCo$_2$Sn with Nb results in a shifting of the Weyl-points closer to the Fermi level. By computing the 2D-ACAR spectra along the [001] direction we obtain the momentum density in the $(p_x,p_y)$ plane and study the consequences of the presence of  Weyl-points in the vicinity of $E_F$.

\section{Density Functional theory calculations}
\label{method}
\subsection{Electronic structure}
Band structures calculations have been performed within the
DFT~\cite{jo.gu.89,kohn.99,jone.15}
using the fully relativistic SP-KKR package \cite{eb.ko.11}.
Both the pure ZrCo$_2$Sn and the Niobium doped Heusler compounds crystallize in the 
face centered cubic symmetry (space group Fm-3m, nr. 225). While the lattice parameter for the pure compound has been experimentally determined to be $a=11.85$ a.u., we considered a lattice parameter of $a=11.77$ a.u.~\cite{wa.ve.16} for both the pure and the Niobium doped Zr$_{0.7}$Nb$_{0.3}$Co$_2$Sn. 
Note that in Ref.~\onlinecite{wa.ve.16}, results for the doped Zr$_{1-x}$Nb$_x$Co$_2$Sn, were presented for a smaller amount of Nb, $x=0.275$. 
For both materials a {\bf k}-mesh of $22 \times 22 \times 22$ points has been used. The general gradient approximation (GGA) for the exchange-correlation energy using the Perdew, Burke and Ernzerhof (PBE) parametrization was applied~\cite{pe.bu.96}. Additionally, the on-site Coulomb interaction from the localized 3d electrons of Co has been accounted for by the GGA+U method~\cite{eb.pe.03} using  $U=3.0$\,eV for the on-site Coulomb interaction and
  $J=0.9$\,eV for the Hund exchange interaction in agreement with Wang et al.~[\onlinecite{wa.ve.16}].    
The local Coulomb interaction in the valence Co $3d$ orbitals were included
via an on-site electron-electron interaction in the form:
$ \frac{1}{2}\sum_{{i \{m, \sigma \} }} U_{mm'm''m'''} c^{\dag}_{im\sigma}c^{\dag}_{im'\sigma'}c_{im'''\sigma'}c_{im''\sigma} $.
Here, $c_{im\sigma}/c^\dagger_{im\sigma}$ annihilates/creates an electron with
spin $\sigma$ on the orbital $m$ at the lattice site $i$.
The Coulomb matrix elements $U_{mm'm''m'''}$ are expressed in the usual
way~\cite{im.fu.98} in terms of Slater integrals.
The double-counting is treated using the so-called atomic limit expression derived by Czyczyk and Sawatsky \cite{cz.sa.94}. Several other double-counting schemes exists, for a more complete discussion the reference Ref.~[\onlinecite{pe.ma.03}] can be consulted. 

The spectral function with or without the spin-orbit coupling was already discussed by Wang et al.~[\onlinecite{wa.ve.16}]. Similarly to results presented in Ref.~[\onlinecite{wa.ve.16}] we find that the majority states around $E_F$ have a dominant Co, respectively Zr $3d-$character. In the GGA calculations the ground state is half-metallic ferromagnetic with an overall magnetic moment in the unit cell of $2\mu_B$. The $3d$-metals in the Heusler compounds have in general weak spin-orbit coupling~\cite{ma.sa.04} which leads
to a slight depolarization. The easy axis magnetization has been found along the [110] direction in agreement with previous calculations~\cite{wa.ve.16}. 
Including the $+U$ correction, both the [110] and the [001] direction for the orientation of the magnetic moment can be stabilized, however, in our calculations, the [001] direction is energetically more favorable. The energy difference amounts to about $4.5 \cdot 10^{-4}$\,Ry. 

\subsection{Positron annihilation spectroscopy}

2D-ACAR is a powerful tool to investigate the bulk electronic structure~\cite{Weber2015,Ceeh2016}.
It is based on the annihilation of positrons with electrons of a sample leading to the emission of
two $\gamma$-quanta in nearly anti-parallel directions. The small angular deviation from
collinearity is caused by the transverse component of the electron's momentum. The coincident
measurement of the annihilation quanta for many annihilation events yields a projection  of the so called two photon momentum density (TPMD) $\rho^{2\gamma}(\bf{p})$.
This is usually computed as the Fourier transform of the product of positron wave function 
$\Psi^+(\bf{x})$ and electron wave function $\Psi^-({\bf{x}})$:
\begin{equation}
	\rho^{2\gamma}({\bf p})\propto \sum_{j,k} n_j({\bf k}) \left| \int d {\bf x} \, e^{-i 2 \pi {\bf x p}} \, \Psi^+({\bf x}) \Psi_{j,{\bf k}}^-({\bf x}) \, \sqrt{\gamma({\bf x})} \right| ^2
    \label{eq:rho2gamma}
\end{equation}
The sum runs over all states $\bf{k}$ in all bands $j$ with the occupation $n_j(\bf{k})$. The so-called ‘enhancement factor’ $\gamma({\bf x})$~\cite{Jarlborg1987}, takes into account the electron positron correlation.  The 2D-ACAR spectrum $N(p_x,p_y)$, the quantity which is actually accessible by an experiment,  is a 2D projection of the 3D momentum-density distribution $\rho^{2\gamma}(\bf{p})$ along a chosen ($p_z$) axis.
\begin{equation}\label{eq:2dacar}
N(p_x,p_y) =  \int \rho^{2\gamma}({\bf p}) dp_z  
\end{equation}

The {2D-ACAR} spectra possess certain symmetries depending on the projection direction. 
The $[100]$ projection in particular possess the symmetry group of a square 
$D_4$\footnote{$D_4$ is the Schoenflis notation 
for a group containing four mirror symmetries in addition to the four fold rotational symmetry}. However, due to the anisotropic resolution function, the symmetry is reduced to the two fold symmetry $D_2$. In order to enhance the statistical accuracy we took advantage of this to get a symmetrized spectrum: 
$	\hat{N}(p_x,p_y) = \sum_{g \in D_2} g[N(p_x,p_y)] $.
The positron annihilation probes all electrons in the system. Filled bands, especially bands of core electrons, give a nearly isotropic distribution which is superimposed by an anisotropy contribution mainly produced by the electrons near the Fermi level. This anisotropic  $A(p_x,p_y)$  contribution is therefore the most interesting feature of an ACAR spectrum $N(p_x,p_y)$. It can be calculated by subtracting isotropic features:
\begin{equation}\label{eq:A}
A(p_x,p_y) = \hat{N}(p_x,p_y) - C(p_x,p_y)
\end{equation}
The radial average  $C(p_x,p_y) \equiv C(\sqrt{p_x^2+p_y^2})$  is constructed from the original spectrum $\hat{N}(p_x,p_y)$ averaging over all data points in equidistant intervals $[p_r,p_r+\Delta p_r)$ from the center.

The DFT can be generalized to electron-positron systems by including the positron density, in
the form of the two-component DFT~\cite{bo.ni.86,pu.ni.94}. In the present calculations
the electron-positron correlations are taken into account by a multiplicative
(enhancement) factor, resulting from the electron-positron interaction included
in the form of an effective one-particle potential as formulated in DFT by Boronski and Nieminen~\cite{bo.ni.86}.

In the LDA(+U) framework the electron-positron momentum density $\rho_{\sigma}({\bf p})$ is computed directly from the two-particle Green function in the momentum representation~\cite{be.ma.06,be.mi.12,ch.be.14}. 
The neglect of electron-positron correlations corresponds to the factorization of the two-particle Green function in real space. In the numerical implementation the position-space integrals for the ``auxiliary'' Green function $G_{\sigma \sigma^{\prime}}({\bf p}_e,{\bf p}_p)$ obtained within LDA or LDA+U, respectively, are performed as integrals over unit cells:
\begin{equation}
\begin{split}
G^{X}_{\sigma \sigma^{\prime}}({\bf p}_e,{\bf p}_p, E_e, E_p) =
\frac{1}{N \Omega}\int d^3{\bf r} \int d^3{\bf r}^{\prime} 
&\phi_{{\bf p}_e \sigma}^{e\dagger}({\bf r}) \,
Im \, G^{X}_{e \  \sigma}({\bf r},{\bf r}^{\prime},E_e) \,
\phi_{{\bf p}_e \sigma}^{e}({\bf r}^{\prime}) \times \\
&\phi_{{\bf p}_p \sigma^{\prime}}^{p\dagger}({\bf r}) \,
Im \, G_{p^+ \ \sigma^{\prime}}({\bf r}, {\bf r}^{\prime},E_p) \,
\phi_{{\bf p}_p \sigma^{\prime}}^{p}({\bf r}\,') \nonumber
\end{split}
\end{equation}
Here $X$ = LDA or LDA+U, and $({\bf p}_e, \sigma)$, and $({\bf p}_p,\sigma^{\prime})$ 
are the momenta and spin of electron and positron, respectively. $G^{X}_{\sigma
\sigma^{\prime}}$ is computed for each energy point on a complex energy contour, providing 
the electron-positron 
momentum density:
\begin{eqnarray} \label{rho_ep}
 \rho_{\sigma}^{X}({\bf p}) &=&  -\frac{1}{\pi} \int dE_e
G^{X}_{\sigma \sigma^{\prime}}({\bf p}_e,{\bf p}_p, E_e, E_p). 
\end{eqnarray}
In Eq.~\ref{rho_ep} integration over positron energies $E_p$ is not required, since only 
the ground state is considered, and enters as a parameter. Moreover in this formalism the positron is considered to be thermalized and described by a state with ${\bf p}_p =0$ with $s$-type symmetry, at the bottom of the positronic band. In addition $\sigma^\prime = -\sigma$ at the annihilation. The 
momentum carried off by the photons is equal to that of the two particles up to a reciprocal
lattice vector, reflecting the fact that the annihilation takes place in a crystal. Hence an
electron with wave vector ${\bf k}$ contributes to $\rho^{X}_{\sigma}(\textbf{p})$ not only 
at ${\bf p} = {\bf k}$ (normal process) but also at ${\bf p} = {\bf k} + {\bf K}$, with 
${\bf K}$ a vector of the reciprocal lattice (Umklapp process). The corresponding 2D-ACAR 
spectrum is computed according to Eq.~(\ref{eq:2dacar}).

\subsection{Signatures of Weyl points in the 2D-ACAR spectra}

\begin{figure}[htp]
\begin{center}
\includegraphics[width=0.35\linewidth]{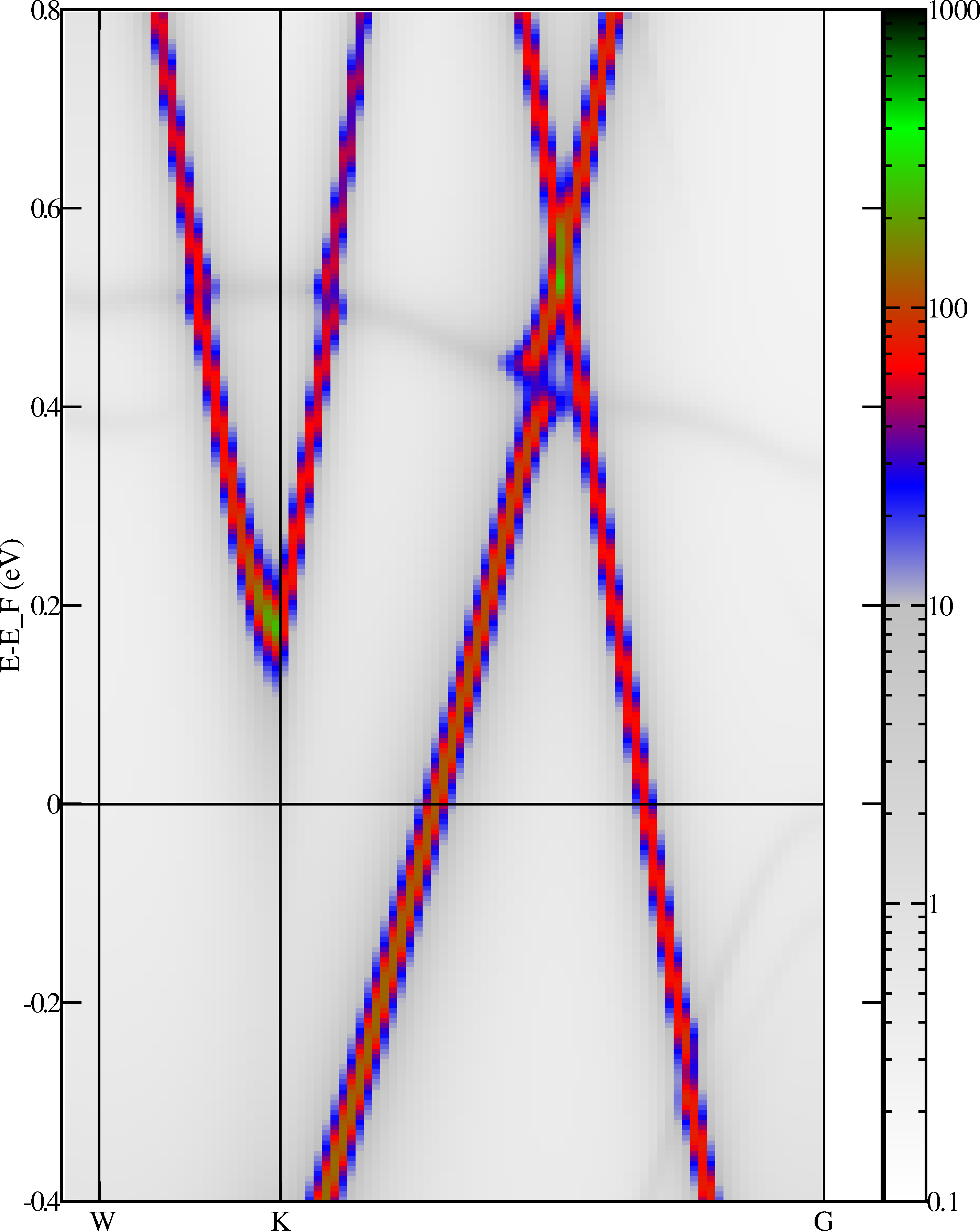}\hspace*{5mm}
\includegraphics[width=0.45\linewidth, clip=true]{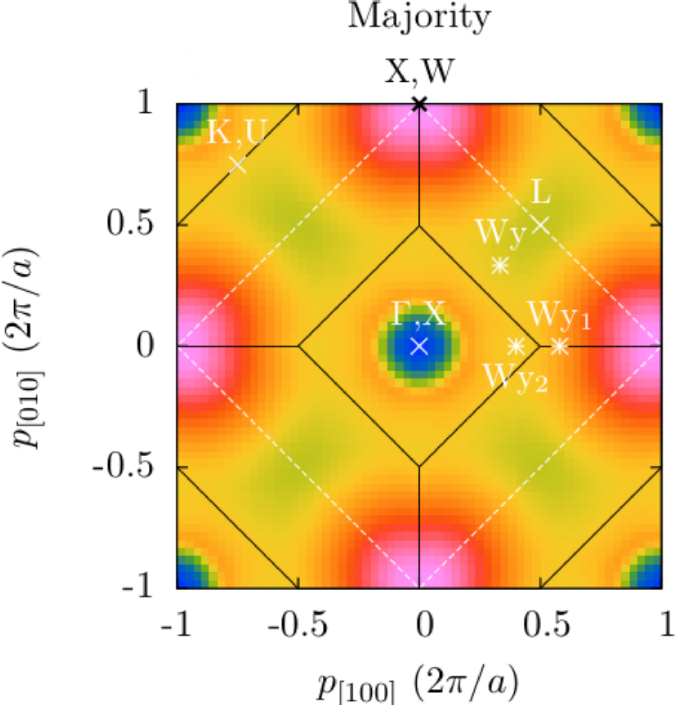}
\includegraphics[width=0.15\linewidth, clip=true]{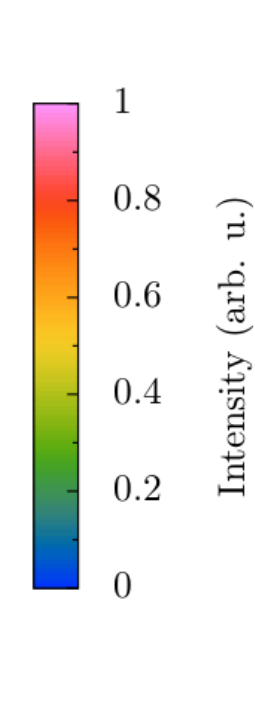}
\end{center}
\caption{Left: Computed Bloch spectral function of ZrCo$_2$Sn, for the majority spin channel, 
which shows the Weyl points situated just above the E$_F$ at $\approx 0.6$\,eV;
Right: LCW back folded momentum density spectra $\rho_{\uparrow}^{GGA+U}({\bf p})$. The solid lines indicate the boundaries 
of the BZ. The color scale encoding the intensity is relative to the value at $N(0,0)$. 
The dashed white line of panel represents the region containing the nodal line.}
\label{fig1:LCW_zrco2sn}
\end{figure}

Information concerning the Fermi surface geometry can be obtained by projecting $\rho^{2\gamma}$
back into the first irreducible Brillouin zone (IBZ). This so-called Lock-Crisp-West (LCW) procedures~\cite{Lock1973} enhances the Fermi surfaces signature as the filled bands yield approximately a constant background.
In Fig.~\ref{fig1:LCW_zrco2sn} we present the spectral function together with the backfolded spectra of pure ZrCo$_2$Sn for the majority spin channel. The results correspond to the GGA+U calculations in which the magnetic moment was oriented along the [110] direction, corresponding to the easy axis magnetisation~\cite{wa.ve.16}. The spectral function is plotted along the $W-K-\Gamma$ direction. 

With the magnetic moment along the [110] direction and in the presence of the spin-orbit coupling the mirror symmetries $M_{x/y/z}$ are broken~\cite{wa.ve.16}. 
We found the nodal $W$ point located in the $ \langle k_x k_y 0 \rangle$-plane at $(0.334, 0.334, 0)$, in units of $2\pi/a$, and in agreement with the work of Wang et. al.~[\onlinecite{wa.ve.16}]. With respect to the Fermi level the Weyl points are situated at $E_F + 0.6$eV. Two other types of Weyls points have been reported in Ref.~[\onlinecite{wa.ve.16}]. However, those appear at slightly different energies in the band structure, and are unstable in contrast to the W-points that are protected by inversion.     
A Weyl point is seen Fig.~\ref{fig1:LCW_zrco2sn} along the $K-\Gamma$ direction, and its pair related by the inversion symmetry is situated along $\Gamma - (-K)$.
Note that within the spectral function a broken line is visible just below the Weyl point, 
and indicates the  mixing with the minority spin channel generated by the full spin-orbit coupling.

The 2D-ACAR spectra is shown on the right hand side of Fig.~\ref{fig1:LCW_zrco2sn}.
The coordinates of three-types of Weyl points (Wy, Wy1, Wy2) have been previously reported Ref.~\onlinecite{wa.ve.16} and are displayed in the 2D-ACAR spectra Fig.~\ref{fig1:LCW_zrco2sn}. 
Wy and Wy1 are situated within the ($p_x,p_y$) plane while Wy2 is out of plane~\cite{wa.ve.16}.
In the center of the Brillouin zone, at the projection of the the $\Gamma$ [0,0,0]-point, a reduced signal intensity is obtained.  The highest signal intensity is observed in the  $X$, [0,2$\pi$/a,0]-point of the IBZ, which coincides with the $W$ [$\pi$/a,2$\pi$/a,0] in the depicted projection. A less intense signal is visible along
the [110]- direction, which goes through the $K$ [3$\pi$/2a,3$\pi$/2a,0]-point, the projection of the $L$ [$\pi$/a,$\pi$/a,$\pi$/a]-point in the ($p_x,p_y$)-plane. Along the 
diagonal in the ($p_x,p_y$)-plane, the light green area, the momentum density 
$\rho_\sigma ({\bf p})$ has a reduced intensity. In Eq.~(\ref{rho_ep}) integration 
is performed until the Fermi level. As the position of the Weyl point is 
situated above $E_F$ the energy range $[E_F, E_F+0.6eV]$ falls out of the integration range.
We have checked that the linear dimension of the light green region, along the 
diagonal corresponds to the opening along the $\Gamma - K$ direction and extending 
the upper limit of the energy integration towards the Weyl point at $E_F+0.6eV$ increases the weight along the $K-\Gamma$ direction. We believe that this constitute an indirect indication of the existence of the Weyl points situated above $E_F$.

\begin{figure}[htp]
\begin{center}
\includegraphics[width=0.40\linewidth, clip=true]{bilder/Fig2_maj_n}
\includegraphics[width=0.40\linewidth, clip=true]{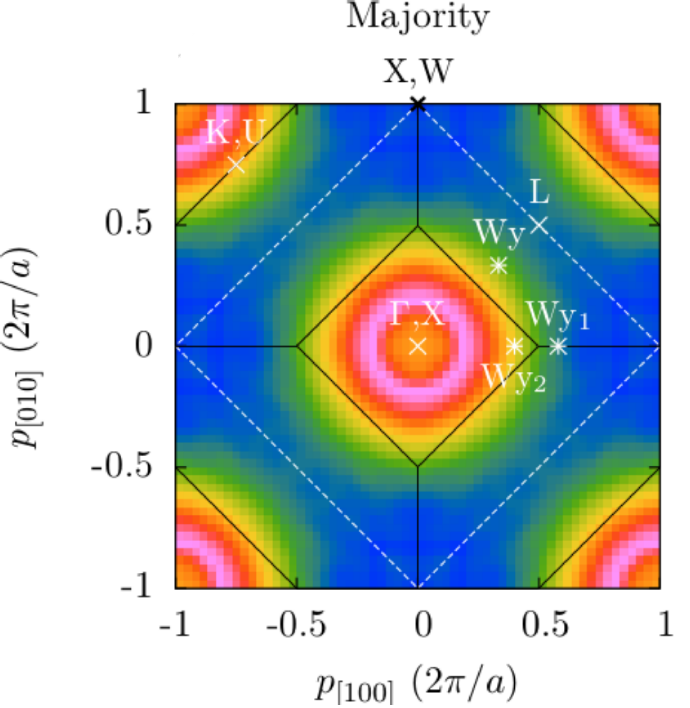}
\includegraphics[width=0.145\linewidth, clip=true]{bilder/Fig_scale}
\end{center}
\caption{Left: LCW back folded momentum density spectra $\rho_{\uparrow}^{GGA+U}({\bf p})$;
Right: back-folded anisotropy spectra $A(p_x,p_y)$ for the majority spin ($\uparrow$) of ZrCo$_2$Sn.}
\label{fig1_2:LCW_zrco2sn}
\end{figure}

In Fig.~\ref{fig1_2:LCW_zrco2sn} we compare the LCW back-folding of the momentum density 
with the back-folded anisotropy spectra.
One observes that for the the majority spins, radial average $C(p_x,p_y)$ removes the spectral weight along the direction connecting $X,W$ points, and suppress completely the high intensities in these points. At the same time the intensity at the $\Gamma$ point is enhanced. No significant change takes place for the minority electrons (not shown). 
The intensity around the projection of the $L$ point into the ($p_x,p_y$)-plane is 
slightly reduced.

The analysis of the dispersion along the nodal line has been presented in Ref.~\cite{wa.ve.16}.
The minimum of the dispersion was found along the $\Gamma - X$ direction, while the maximum is realized along the $\Gamma - K$ direction. The spin-orbit orbit coupling would gap the nodal lines, however along the [110]-direction the Weyl points survive being protected by a $C_2$ symmetry along the [110]-direction. 
This analysis is supported by the anisotropy plot, the momentum density is depleted along the 
$\Gamma-X$ while only a smaller reduction is seen along the $\Gamma-K$ direction.
%

%

The Weyl point situated at $E_F+0.6$eV, in the clean compound ZrCo$_2$Sn 
can be brought to the Fermi level by a proper concentration of alloying as discussed
by {\it Wang et. al.}~\cite{wa.ve.16}. Similarly, we also performed the 
electronic structure calculations for the Nb$_{0.3}$Zr$_{0.7}$Co$_2$Sn alloy 
using the Coherent Potential Approximation (CPA)~\cite{sove.67,gyor.72} as implemented in the KKR~\cite{eber.00,eb.ko.11}. As {\bf k} is not a good quantum number in disordered alloys, one can still define within the CPA an effective Fermi surface. Due to disorder smearing of the Fermi surface, the magnitude of this effect represents however a small fraction of the FS dimensions.  

\begin{figure}[h]
\centering
\includegraphics[width=0.35\linewidth]{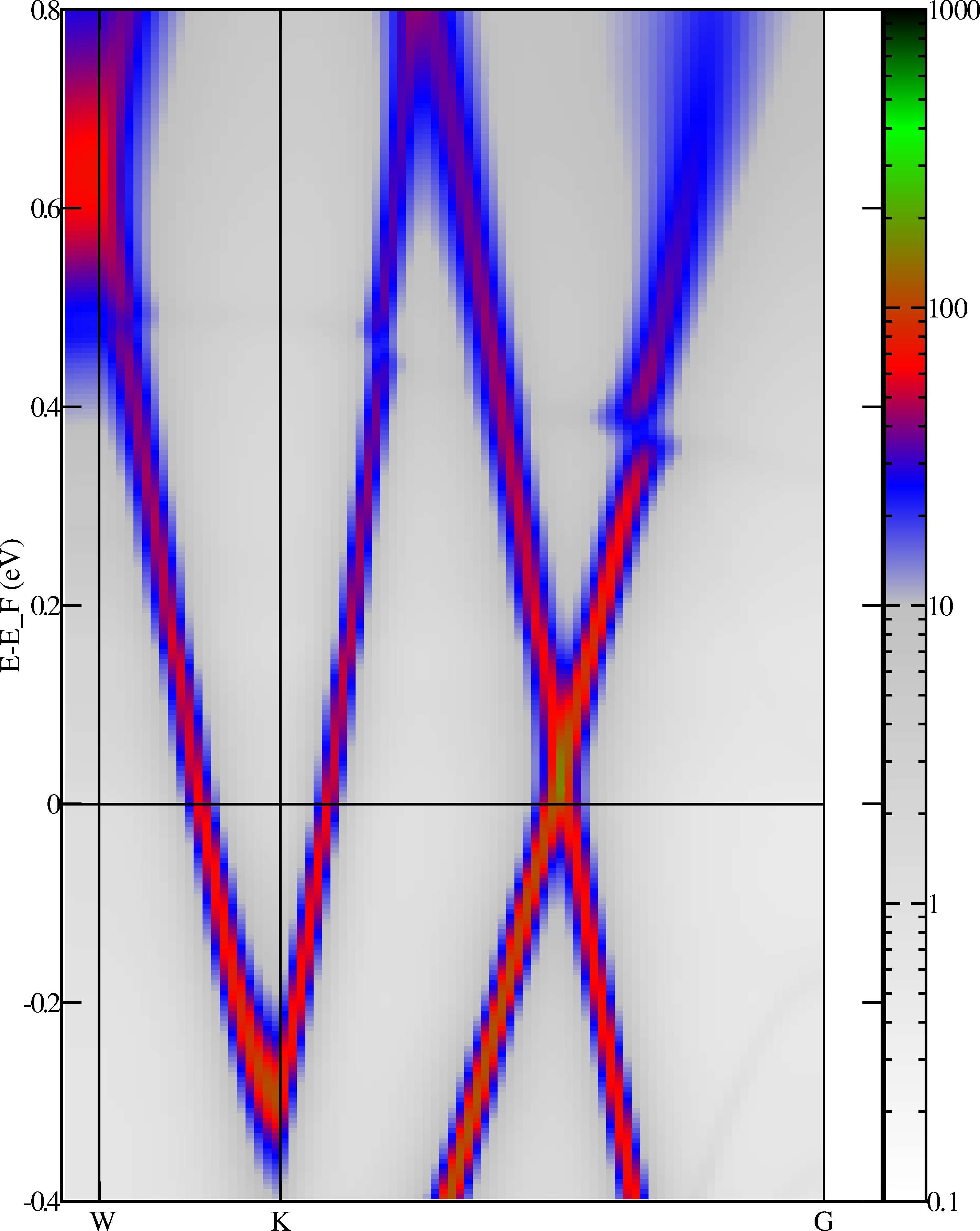}\hspace*{5mm}
\includegraphics[width=0.45\linewidth, clip=true]{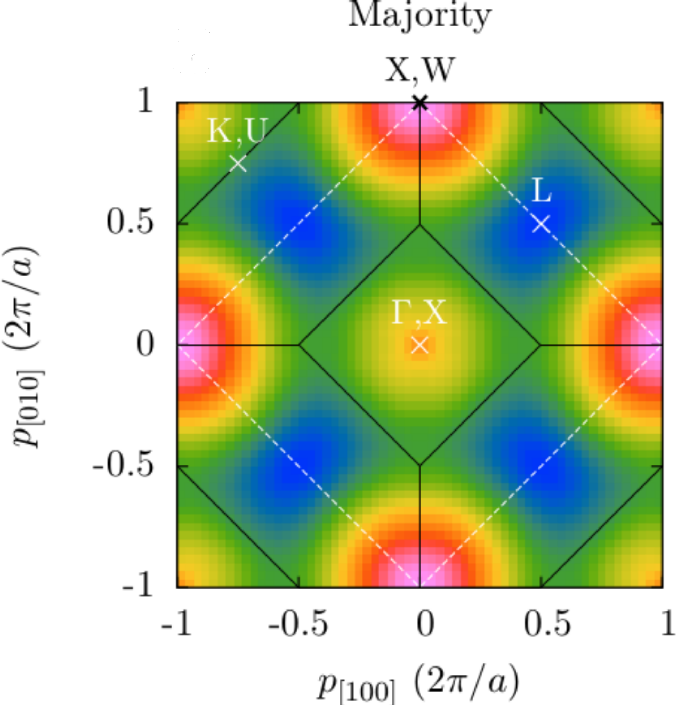}
\includegraphics[width=0.15\linewidth, clip=true]{bilder/Fig_scale}
\caption{Left: The spectral function for Zr$_{0.7}$Nb$_{0.3}$Co$_2$Sn, the Weyl point is close to the Fermi level. 
Note that disorder effects broaden the spectral function lines.
Right: Back-folded anisotropy spectra $A(p_x,p_y)$ of ZrCo$_2$Sn computed within the GGA+U. The solid lines indicate the outline of the BZ.  
}
\label{fig2:LCW_zrco2sn}
\end{figure}

On the left part of Fig.~\ref{fig2:LCW_zrco2sn} we present the spectral function for the disordered alloy. By alloying the most important features of the Fermi surface remain the 
same, as expected because the main band topology is unchanged~\cite{wa.ve.16}. 
The same symmetry analysis holds for the Niobium doped Zr$_{0.7}$Nb$_{0.3}$Co$_2$Sn system as in the undoped case. 
The spin-orbit induced feature visible at about $E_F+0.4eV$, is present also in the case of disorder as expected, its position remain unchanged, while the Weyl point is brought closer to $E_F$.
As the Weyl-point approaches $E_F$ the entire manifold of bands are almost rigidly shifted towards lower energies.

The complex coherent potential contributes by smearing out the Green's function
and consequently the features in momentum space. By construction the coherent potential is local (no momentum dependence), therefore it's smearing is isotropic. By back-folding the  anisotropy spectra we eliminate this effect. 
The anisotropy is computed according to Eq.~\ref{eq:A} and is seen on the right hand side 
of Fig.~\ref{fig2:LCW_zrco2sn}. 
As expected, moving the Weyl point towards $E_F$, leads to a depletion of momentum density. However the 2D-ACAR spectra does not allow for a more quantitative analysis, a more detailed discussion is presented below.

\section{Discussion and Conclusion}
Weyl semi-metals is a new class of metallic phases that contains Weyl points (i.e. crossing
points of two non-degenerate bands) in the band structure near the Fermi level. 
Proposed experiments to identify Weyl semi-metals exist in the literature: magnetoresistance~\cite{aji.12,so.sp.13}, coupling between collective modes~\cite{ch.pe.13}, and transport~\cite{pa.gr.12} measurements. 
The purpose of this paper is to investigate 
theoretically the momentum density and to propose 2D-ACAR measurements as a systematic 
way to examine the existence of the Weyl points and their behavior in certain situations such as disorder. 

We studied the pure Heusler ZrCo$_2$Sb and the Niobium doped NbZrCo$_2$Sb, in which Weyl nodes are not too far from the Fermi energy. 
We have performed ab-initio calculations for the spectral functions and for the 2D-ACAR spectra for both the pure and the alloyed compound. Our band structure calculations 
confirm the existence of two Weyl points related by inversion symmetry situated along 
the [110] direction, as reported previously~\cite{wa.ve.16}. By doping with Nb the Weyl point move closer to $E_F$ which is visible in the spectral function. 
From our calculations for the 2D-ACAR spectra we can not undoubtedly identify the position 
of the Weyl points in the pure, nor in the doped, Heuslers. Our numerical analysis is conducted by comparing results of the energy integration  Eq.~(\ref{rho_ep}) with different upper bounds.
In all results, a relatively large smearing at 
the Fermi surface in the angular correlation curve is obtained, which hinder the quantitative evaluation for the position of the Weyl-points. The current qualitative analysis can not totally attribute this difficulties to positron wave function effects, an additional important 
effect that is not considered here is the direct electron-positron Coulomb interaction.
Early calculations~\cite{kaha.63,ca.ka.65,carb.67}, taking the Coulomb interactions 
into consideration in the simplified picture of the electron gas were able to 
explain the small deviations for $p < p_F$. However the theoretical results~\cite{kaha.63,ca.ka.65,carb.67} for
momenta beyond $p_F$, fail to explain measured features, because  
the cancellation effects among the diagrams in perturbation expansion in 
the electron-positron interaction.  
Therefore, to resolve the Weyl features occurring above $E_F$ an approach beyond the perturbative analysis~\cite{kaha.63,ca.ka.65,carb.67} is required. However, the current DFT, GGA(+U) calculations limit themselves to the independent particle model, which may well be the reason why the Weyl points are not resolved in the momentum density spectra. 
In recent studies we partly include dynamical Coulomb interactions effects upon the 2D-ACAR spectra at least for the electronic subsystem~\cite{Ceeh2016,we.be.17}. Further work along the lines to include the direct electron-positron interaction is in progress.
The current methodology, may be nevertheless relevant for another ferromagnetic full Heusler compound Co$_2$MnGa~\cite{gu.su.17} with Weyl points, just below the Fermi level.

\section*{Acknowledgements}
This project is funded by the Deutsche Forschungsgemeinschaft (DFG) within the Transregional Collaborative Research Center TRR 80 ``From electronic  correlations to functionality''. D.B. acknowledges financial support provided through UEFISCDI grant PN-II-RU-TE-2014-4-0009 (HEUSPIN). We would like to thank H. Ebert and J. Min\'ar for the fruitful collaboration.

\bibliography{references.bib}
\end{document}